\documentclass[twocolumn, switch]{article} 

\usepackage{preprint}
\usepackage{fancyhdr,graphicx,amsmath,amssymb}
\usepackage[ruled,vlined]{algorithm2e}
\usepackage{tablefootnote}

\usepackage{amsmath, amsthm, amssymb, amsfonts}

\usepackage[numbers,square]{natbib}
\usepackage[utf8]{inputenc}	
\usepackage[T1]{fontenc}	
\usepackage{xcolor}		
\usepackage[colorlinks = true,
            linkcolor = purple,
            urlcolor  = blue,
            citecolor = cyan,
            anchorcolor = black]{hyperref}	
\usepackage{booktabs} 		
\usepackage{nicefrac}		
\usepackage{microtype}		
\usepackage{lineno}		
\usepackage{float}			

\usepackage{lipsum}		

\usepackage{newfloat}
\DeclareFloatingEnvironment[name={Supplementary Figure}]{suppfigure}
\usepackage{sidecap}
\sidecaptionvpos{figure}{c}

\usepackage{titlesec}
\titlespacing\section{0pt}{12pt plus 3pt minus 3pt}{1pt plus 1pt minus 1pt}
\titlespacing\subsection{0pt}{10pt plus 3pt minus 3pt}{1pt plus 1pt minus 1pt}
\titlespacing\subsubsection{0pt}{8pt plus 3pt minus 3pt}{1pt plus 1pt minus 1pt}

\usepackage{tikz,xcolor}
\definecolor{lime}{HTML}{A6CE39}
\DeclareRobustCommand{\orcidicon}{
	\begin{tikzpicture}
	\draw[lime, fill=lime] (0,0) 
	circle [radius=0.16] 
	node[white] {{\fontfamily{qag}\selectfont \tiny ID}};
	\draw[white, fill=white] (-0.0625,0.095) 
	circle [radius=0.007];
	\end{tikzpicture}
	\hspace{-2mm}
}
\foreach \x in {A, ..., Z}{\expandafter\xdef\csname orcid\x\endcsname{\noexpand\href{https://orcid.org/\csname orcidauthor\x\endcsname}
			{\noexpand\orcidicon}}
}

\title{StrokeSight: A Novel EEG-Based Diagnostic System for Strokes Using Spectral Analysis and Deep Learning

}

\usepackage{authblk}

\author[1,2]{Rohan Kalahasty}
\author[1,2]{Lakshmi Sritan Motati}

\affil[1]{Harvard Medical School}
\affil[2]{Thomas Jefferson High School for Science and Technology}

\begin{document}

\twocolumn[ 
  \begin{@twocolumnfalse} 
  
\maketitle

\begin{abstract}
A stroke is defined as a neurologic deficit arising from an interruption in blood supply to the brain. According to the World Health Organization, over 15 million people suffer from strokes annually, of which almost 70\% die or are permanently disabled. Effective treatment must be administered within one hour to prevent irreversible brain damage. Unfortunately, the current gold standards for diagnosis, CT and MRI, are time-consuming, expensive, and immobile. Electroencephalograms reveal biomarkers of strokes while being inexpensive and available for remote use, but no system exists that utilizes them for this purpose. To address this issue, we created StrokeSight, a novel, open-source web application that automatically provides a full diagnosis and visualization of ischemic and hemorrhagic strokes in under 50 seconds using 60-second electroencephalograms. We first calculated the averaged power spectral densities for 132, 60-second electroencephalogram readings, which we then used to train three deep neural networks that respectively predict a stroke’s type (control/ischemic/hemorrhagic), location (left/right hemisphere), and severity (small/large) with accuracies of 97.5\%, 94.4\%, and >99\%. StrokeSight also implements a novel process to visualize spectral abnormalities caused by strokes. Azimuthal equidistant projection and multivariate spline interpolation are used to reshape 3D electrodes onto a head-shaped 2D plane and then a contour map of each frequency band power is created, allowing neurologists to quickly and accurately interpret electroencephalogram data. StrokeSight could act as a revolutionary solution for stroke care that drastically improves the speed, cost efficiency, and accessibility of stroke diagnosis while allowing for personalized treatment and interpretation.
\end{abstract}
\vspace{0.35cm}

  \end{@twocolumnfalse} 
] 



\section{Introduction}
Strokes are caused by a clot or burst in a blood vessel that carries nutrients to brain tissue \cite{strokereview}. The effects of a stroke are mostly irreversible if a stroke is not treated within the first 60 minutes and can cause impairment in speech, language, cognitive skills, and vision \cite{lempriere_2021}. The current gold standards for stroke diagnosis are computerized tomography (CT) and magnetic resonance imaging (MRI). However, these two imaging modalities present a plethora of issues. MRI and CT scans, respectively, expose patients to high levels of ionizing radiation and magnetic fields, which raises concerns about potential side effects \cite{ABUELHIA2020295} \cite{franklen}. Furthermore, these scans are highly expensive for both patients (\$1,600 to \$8,400) \cite{mricost} and clinics, which propagates large disparities in stroke mortality rates among minority groups including Black Americans and American Indians compared to White Americans \cite{strokeracism}. MRI and CT scans are also not portable and can take up to an hour to get. On top of accessibility issues, the accuracy of these scans is unreliable, as they rely on qualitative analysis by professionals. For example, four neurologists came to a unanimous diagnosis regarding stroke only 58\% of the time with a CT scan and 80\% of the time with an MRI scan \cite{chalela}. These issues make CT and MRI scans an inadequate method to act as the gold standard for stroke diagnosis, and a new method is needed to allow for faster diagnosis in both remote and clinical locations. 
\\
\\
Electroencephalography (EEG), is a testing modality for brain function that measures electrical activity from electrodes placed on multiple regions of the scalp. Electroencephalogram technology has been developing rapidly in the past years because they are being applied to fields outside of medicine, such as brain-computer interfaces (BCI) \cite{powers_2021}. As a result, EEG headsets now come in a variety of different form factors, and many are highly portable, easy to use, and allow for remote monitoring \cite{10.3389/fninf.2020.553352}. EEGs have been shown to contain many biomarkers of strokes that can be used as a diagnostic basis. It has been shown that people who have suffered a stroke display reduced cortical activity and connectivity in both the alpha and beta waves and an increase in activity in the gamma wave \cite{snyder_schmit_hyngstrom_beardsley_2021}. These abnormalities are driven by the drop in cerebral blood flow to brain tissue during a stroke. Furthermore, asymmetries in activity will likely be seen as a reduction of activity in the affected hemisphere \cite{snyder_schmit_hyngstrom_beardsley_2021}. EEGs could thus be a powerful method for fast and remote stroke diagnosis, but no system currently exists that leverages EEGs to return a full personalized patient diagnosis. StrokeSight bridges this gap by proposing a web application-based stroke diagnostic framework that can take in a 60-second EEG recording and return a personalized diagnosis and visualizations of brain activity. 

\section{Materials and Methods}
\subsection{Dataset Description}
The dataset we used to train our machine learning models was prepared by Goren et al. and the Hyper Acute Stroke Unit at University College London Hospital (UCLH) \cite{Goren2018}. This dataset contains multi-frequency electrical impedance tomography (MFEIT) data, which was collected by passing electrical currents through surface electrodes at a large range of frequencies. The hardware used included a BioSemi EEG Recorder, a 32 electrode setup of EasyCap EEG electrodes, and various custom components. This system recorded frequencies from 5 Hz to 2 kHz with impedance measurements at various frequencies. Among the files in the dataset, we used the time difference (TD) files for each patient; these recordings were collected at 200 Hz, 1.2 kHz, and 2 kHz, for a total of 60 frames collected over 25 minutes. The recordings were collected in 10 healthy volunteers, 10 patients with ischemic stroke, eight patients with hemorrhagic stroke, and five patients with other classifications of stroke (these five were not used in our study). Some patients had multiple recordings, leading to a total of 44 usable readings. Among these readings, 20 were from healthy subjects, 14 were from ischemic stroke patients, and 10 were from hemorrhagic stroke patients. The majority of the data was collected within 24 hours of the onset of the stroke. Additionally, the dataset contained annotations and radiology reports which gave information such as the size (small or large) and location (left or right hemisphere) of the stroke. Among the 24 readings from patients that suffered from a stroke, the split of small and large strokes (severity) was 8 to 16, respectively. The split of strokes occurring in the left and right hemispheres of the brain in the same readings was 10 to 14, respectively.

\subsection{Preprocessing}
Within the EIT recordings, there is hidden EEG data that is mixed with electrical interruptions at various frequencies. Thus, to extract only the EEG data, we filtered out certain frequencies. 
We used a bandpass filter that only preserved data from frequencies up to 60 Hz and removed the rest, which contained the non-EEG impedance data (the lowest frequency of the injections was 200 Hz in the Time Difference recordings). We then cropped the recording to only the first 180 seconds and split that into three, equal-sized segments to create a larger dataset for our machine learning models to use for training and testing. EEG's are a nonlinear dynamical systems, each segment of an EEG will not resemble to one before. All of the preprocessing was done with the MNE package in Python \cite{mne}.
\subsection{Feature Extraction}
The feature extraction method chosen was power spectral density because it efficiently condenses data while preserving spatial data. Welch's method was used to estimate the power spectral density (PSD) \cite{welch}. PSD has been shown to extract important features for deep neural network models for EEG classification and biomarker discovery in clinical literature \cite{hasan, perez, rahman, wang}.
\\
\\
Welch's Method for PSD calculation is composed of three major steps. First, the overall time signal is split into successive segments of the signal. The formula for the $m$th segment (frame) of a signal $x$ is shown below, where $w(n)$ is the rectangular window, $R$ is the window hop size (overlap), and $K$ is the number of segments.
\begin{equation}
     x_m(n) = w(n)x(n+mR), \quad n=0,\ldots,M-1,\; m=0,\ldots,K-1,
\end{equation}
Next, the periodogram is calculated for each segment. The periodogram converts the segment from the time domain to the frequency domain, which estimates the power of each frequency range. This is calculated using the fast Fourier transform. The formula for the calculation of the periodogram for a segment $m$ is shown below. 
\begin{equation}
  P_{x_m,M}(\omega_k)
= \frac{1}{M}\left\vert\hbox{\sc FFT}_{N,k}(x_m)\right\vert^2
= \frac{1}{M}\left\vert\sum_{n=0}^{N-1} x_m(n) e^{-j2\pi nk/N}\right\vert^2
\protect
\end{equation}
Finally, the periodograms of all segments are averaged to create a final periodogram for a signal according to the equation.
\begin{equation}
    \displaystyle {\hat S}_x^W(\omega_k) = \frac{1}{K}\sum_{m=0}^{K-1}P_{x_m,M}(\omega_k). \protect
\end{equation}
Through these steps, we estimate the average periodogram over time for one signal, and since every one of the 32 electrodes records its own signal, the power spectral density of the EEG is a two-dimensional matrix, with a 1D averaged periodogram with the powers of various frequency ranges for each electrode. The resultant PSD matrix for an EEG can be represented as:

\begin{equation*}
\begin{matrix}
 & channel\ 1\\
 & \vdots \\
 & channel\ 32
\end{matrix} \ \ \begin{bmatrix}
P_{1,\ 1} \  & \cdots  & P_{1,\ 9}\\
\vdots  & \ddots  & \vdots \\
P_{32,\ 1} & \cdots  & P_{32,\ 9}
\end{bmatrix} \ \ \ \ 
\end{equation*}
\begin{center}
 \ \ \ \ \ \ \ \ \ \ \ \ \ \ \ \ \ \ \ $\displaystyle \ \ \ \ \begin{matrix}
f_{0} & \ \ \dotsc  & \ \ \ \ f_{9}
\end{matrix}$
\end{center}

where the frequency range of $f_n$ is $[4n, 4n + 4)$.

\subsection{Deep Learning}

To use the power spectral density matrices for classification, artificial neural networks are used (ANNs). ANNs have been used for EEG classification in recent years once features have been extracted \cite{gu, maksimenko, tomasevic}. This is because they are able to learn complex, nonlinear relationships in multivariate data. Thus, we hypothesized that using each number in the PSD matrix as a separate feature (since they all represent the activity of a certain frequency at a certain electrode) would allow for the accurate classification of a PSD matrix.

\begin{figure}[H]
    \includegraphics[width=0.48\textwidth]{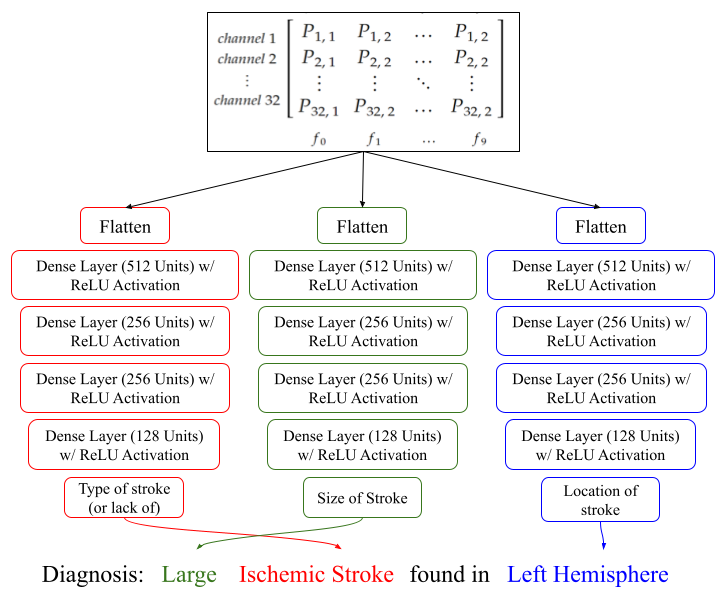}
    \caption{Three ANNs are developed side-by-side, each performing a different task to output a personalized diagnosis.}
    \label{fig:mesh1}
\end{figure}
In this study, we created three ANN models for the detection of strokes (healthy, ischemic, or hemorrhagic), classification of severity (small or large), and localization of a stroke (left or right hemisphere of the brain). The structure of these three models is similar (Figure \ref{fig:mesh1}). First, we used a Flatten layer to convert the 2D PSD matrix into a 1D array of features that is passed into an input layer. Then, data is passed through a series of dense layers each followed by a Rectified Linear Unit (ReLU) activation function. The sizes of the hidden dense layers are as follows: 512, 256, 256, and 128 units. Lastly, the data is passed into the output layer. In the multi-class model for stroke detection, the output layer has three nodes and uses a softmax activation function. In the other two models, the output layer has only one node and uses a sigmoid activation function.
\\
\\
The stroke detection model used the categorical cross entropy loss function, while the other two models used the binary cross entropy (BCE) loss function.
\\
\\
To account for class imbalance in the size and location models, class weighting is used. Class weights are chosen using the conventional manner defined in the following equation 
\begin{equation}
    w_0 = \frac{n_0+n_1}{2n_0} \hspace{10pt}              
    w_1 = \frac{n_0+n_1}{2n_1}
\end{equation}

The cross entropy loss function with these weightings is
\begin{equation}
    L = -[w_1ylogp+w_0(1-y)log(1-p)]
\end{equation}
All models used the Adam optimizer \cite{adamoptimizer}. After training, the three models were saved as H5 files.
\begin{figure*}[h]
    \centering
    \includegraphics[width=.9\textwidth]{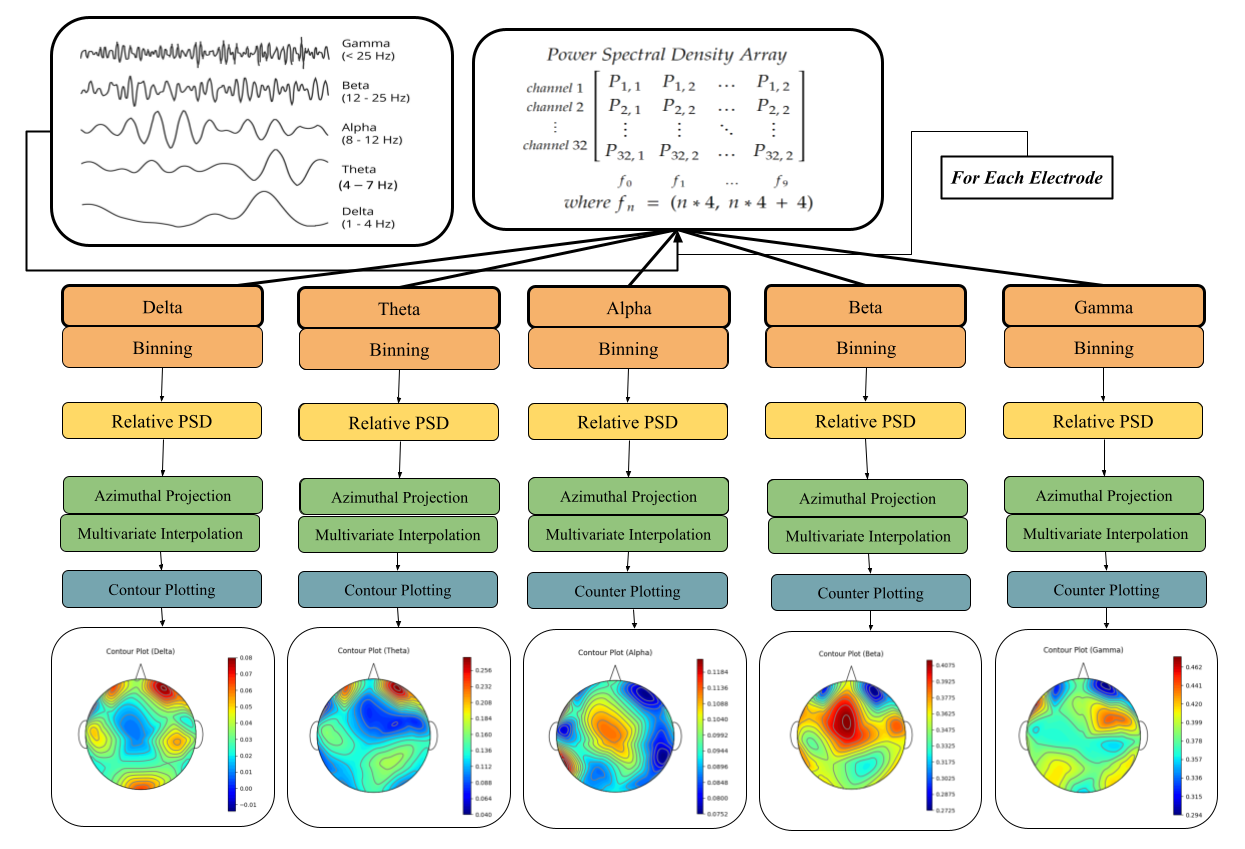}
    \caption{Full pipeline of the developed power spectral density visualization process.}
    \label{fig:viz}
\end{figure*}
\subsection{Visualization Procedure}
In addition to the machine learning models, we created and evaluated a novel visualization process using the power spectral density matrices to display the activity of various brain waves (Figure \ref{fig:viz}). The brain waves, or frequency bands, that we visualize are the delta, theta, alpha, beta, and gamma waves; these correspond to the following wavelength ranges: 1-4 Hz, 4-7 Hz, 8-12 Hz, 12-25 Hz, and >25 Hz (respectively).

First, we find the relative contributions of each frequency to the overall band. This is calculated by dividing the PSD of a frequency in a band by the sum of the PSD of the entire band. This process is represented by the following equation. 
\begin{equation}
    P_{relative} = \frac{\sum^{f=f_2}_{f=f_1}P(f)}{\sum^{f=f_H}_{f=f_L}P(f)}
\end{equation}
where $[f_{L}, f_{H}] = [0, 40]$ and $[f_{1}, f_{2}]$ is determined by the frequency of the band selected (gamma, beta, alpha, theta, and delta.
This is repeated for each electrode, which generates the data for the visualization of each brain wave. Next, we convert the coordinates for the 32 electrodes given in the dataset from a brain-shaped 3D space to a circle-shaped 2D plot. The azimuthal equidistant projection (AEP) is used to project the 3D coordinates onto a 2D plane: the 3D rectangular coordinates are converted to 2D polar coordinates with modified radii depending on the z-coordinates, which gets the 2D rectangular coordinates of the projection. Spline interpolation is then used to reshape the 2D coordinates into a circle shape and normalize the values. Finally, with the relative PSD data and reshaped electrode coordinates, we create a contour plot for each of the five frequency bands. Contour lines, a circular outline for the head, and a nose and ears are also added to the plot to add directionality and interpretability. 
\begin{figure}[H]
    \includegraphics[width=0.5\textwidth]{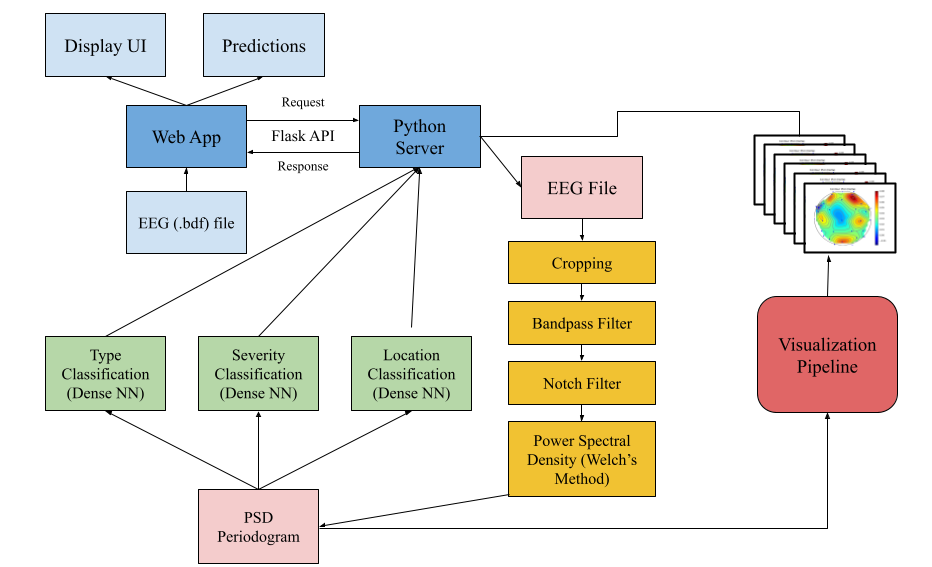}
    \caption{Architecture of the web application for StrokeSight.}
    \label{fig:web}
\end{figure}
\subsection{Web Application}
We developed a web application for StrokeSight that takes an EEG file in the BioSemi data format (.bdf) and uses the saved machine learning models (stored as H5 files) and the visualization algorithms to prepare a personalized diagnosis for a stroke patient.

The app architecture is shown in Figure \ref{fig:web}. We used the Flask framework to create an API that connects the Python backend to the frontend built with HTML, CSS, and JavaScript. A form page takes in the upload of an EEG file, which is downloaded in the backend and processed to retrieve the predictions, confidence percentages for those predictions, and visualizations.
\subsection{Performance Metrics} \label{sssec:performance}
To assess the classification performances for the three developed machine learning models, we used the following statistical measures: accuracy, $F_1$ score, precision, recall, and area under the receiver operating characteristic (AUC). AUC was not calculated for the stroke detection model because it predicted more than two classes. The metrics were calculated as follows (excluding AUC):
\begin{equation}
    accuracy=\frac{TP+TN}{TP+TN+FP+FN}
\end{equation}
\begin{equation}
    F_1=\frac{2*precision*recall}{precision+recall}
\end{equation}
\begin{equation}
    precision=\frac{TP}{TP+FP}
\end{equation}
\begin{equation}
    recall=\frac{TP}{TP+FN}
\end{equation}
TP represents the true positives, which means a data point with a positive label was predicted correctly. TN represents the true negatives, which means a data point with a negative label was correctly predicted to be negative. Similarly, FP (false positive) and FN (false negative) represent data points that were predicted to be positive but are negative (FP) or were predicted to be negative but are positive (FN). In the context of the severity and location prediction models, which are binary classifiers, positive and negative labels can represent the two classes. In the context of the stroke detection model, TP, FP, TN, and FN represent these values for each class summed over every class (ex. TP would be the sum of the TP of "no stroke" vs. the other classes, TP of "ischemic" vs. the others, and TP of "hemorrhagic" vs. the others).

The AUC was calculated for the binary classification models by plotting the receiver operating curve, which involves graphing the false positive rate vs. the true negative rate at various thresholds. A higher AUC score indicates that a model is better at distinguishing between two classes.

\section{Results and Discussion}
\begin{table*}
	\caption{Comparison of our stroke detection model to existing literature *\\}
	\centering
	\begin{tabular}{llllll}
		\toprule
        \bf Model & \bf Accuracy &  \bf F1 Score & \bf Precision & \bf{Recall} & \bf EEG Length (min.) \\
         \bf Proposed Model & \bf 0.95 & \bf 0.975 & \bf 0.975 & \bf 0.975 & \bf 1 \\
         Giri et al. \cite{giri} & 0.861 & 0.861 & 0.870 & 0.861 & 15 \\
         Qureshi et al. \cite{qureshi} & 0.95 & - & - & - & 15-240 \\
		\bottomrule
	\end{tabular}
	\label{tab:table1}
\end{table*}

\begin{table*}
	\caption{Comparison of our stroke severity classification model to existing literature.}
	\centering
	\begin{tabular}{llllll}
		\toprule
        \bf Model & \bf Accuracy &  \bf F1 Score & \bf Precision & \bf{Recall} & \bf AUC \\
         \bf Proposed Model & \bf 1.000 & \bf 1.000 & \bf 1.000 & \bf 1.000 & \bf 1.000 \\
         Wilkinson et al. \cite{wilkinson} & 0.76 & - & - & 0.63 & - \\
		\bottomrule
	\end{tabular}
	\label{tab:table2}
\end{table*}

\begin{table*}
	\caption{Obtained metrics of our stroke localization (hemisphere prediction) model. \\}
	\centering
	\begin{tabular}{llllll}
		\toprule
        \bf Model & \bf Accuracy &  \bf F1 Score & \bf Precision & \bf{Recall} & \bf AUC \\
         \bf Proposed Model & \bf 0.944 & \bf 0.93 & \bf 0.875 & \bf 1.000 & \bf 0.974 \\
		\bottomrule
	\end{tabular}
	\label{tab:table3}
\end{table*}
Tables \ref{tab:table1}-\ref{tab:table3} show the results of the three machine learning models using the metrics described in Section \ref{sssec:performance}. Additionally, these tables compare our obtained results to the metrics of papers that perform the same task.
\\
\\
From the results, we see that all of our models achieved high performance on the testing dataset, which consisted of 20\% of the entire dataset of PSD matrices. With the stroke detection and severity classification models, all of the metrics we achieved were state-of-the-art and were similar to each other. Furthermore, the stroke detection model developed in this study is the only one that performs multi-class classification, between Ischaemic, Hemorrhagic, and no stroke EEGs. As for the localization model, this study is the only existing literature that does this task with EEGs and achieves a high accuracy, F1 score, recall, and AUC, although the precision is visibly lower than the other metrics.
\\
\\
Table \ref{tab:table4} shows the mean speeds of the web application for computing the visualizations and ML-powered predictions once an EEG file is uploaded (the time for uploading the file was not included). This data was collected by measuring the time that each step took 30 times. This application was able to output a full, personalized diagnosis with visualizations for interpretation in under one minute. The ML models took barely any time because we use trained models saved in the H5 format. On the other hand, the visualization likely takes much more time because it includes creating plots and saving them as images locally, which are referenced in the frontend.

\begin{table}[H]
 \caption{Mean speeds of the web application we created over 30 samples.}
  \centering
  \begin{tabular}{ll}
    \toprule
    Component     & Average Time (s)\\
    \midrule
    ML Diagnosis & 1.14 \\
    Visualization & 42.95 \\
    Total & 44.09 \\
    \bottomrule
  \end{tabular}
  \label{tab:table4}
\end{table}

\begin{figure}[H]
    \includegraphics[width=0.5\textwidth]{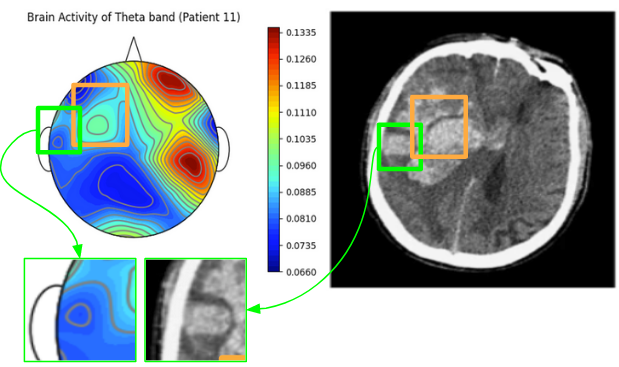}
    \caption{Comparison of visualization of theta wave activity in patient from dataset to the corresponding CT scan.}
    \label{fig:vizeval}
\end{figure}

Finally, we turn to our novel visualization process to evaluate how useful these plots could potentially be to neurologists using EEG technology for stroke assessment. Although this evaluation is more qualitative, this can be viewed as a proof-of-concept for PSD-based visualization of brain activity, which could be used to find visual/spatial biomarkers in diseases other than strokes.\\\\In Figure \ref{fig:vizeval}, we see the theta wave visualization of Patient 11 in the dataset and the corresponding CT scan of the patient. In the plot, we see red spots on the right side of the hemisphere; this behavior is expected, as theta waves are normally more active on the right side of the brain \cite{morillon}. On the left side, however, the color is mostly dark blue, thus indicating that there is very little theta activity. That being said, there is one local peak (highlighted in orange) in the front-left of the brain. This peak is the location of the stroke, and this is clearly visible in the orange region of the CT scan. Additionally, a small cup-shaped abnormality seen in the CT scan (highlighted in green) is also somewhat visible in the visualization. This shows that the visualization process proposed in this study, although not perfect, has the potential to become useful for interpreting EEG readings in a variety of diseases with EEG biomarkers.
\\\\
StrokeSight represents the world's first fast, accurate, and accessible stroke diagnostic system. The results show that we achieve state-of-the-art metrics while using shorter EEG readings. This is due to the ability of power spectral density (PSD) to condense the EEG data while retaining the information that helps with classification. This study validates the use of EEG for making stroke diagnosis more personalized and accurate. Additionally, the novel visualization process that we propose is relatively lightweight and preserves important biomarkers while allowing for the discovery of new biomarkers. These visual cues make the interpretation of EEGs much easier and more accessible, thus making this modality a potential alternative to the currently used MRI and CT scans. Altogether, the StrokeSight system provides the necessary start for EEG-based stroke diagnostics to arise. We show that PSD is a strong feature extraction method for the EEGs of stroke patients, EEG readings can be made interpretable with PSD, and a software-based diagnostic system can be both accurate and fast.

\section{Conclusion}
In this paper, we present a strong proof-of-concept for an EEG-based diagnostic for strokes. StrokeSight is more accurate than neurologists using CT and MRI scans for stroke detection and personalized diagnosis. Furthermore, since EEGs can be taken in remote locations, it also represents the first remote and personalized stroke diagnostic system. The visualization process synthesizes images that are rich in information and can be used to analyze other diseases with power spectral changes as well, such as Alzheimer's disease and Parkinson's disease. This work shows that EEGs can act as a valid diagnostic basis for strokes when implemented with an external system. They have the potential to speed up stroke diagnosis and improve patient outcomes. In the future, the system should be integrated within EEG headsets to allow for real-world diagnosis. In the future, the prospect of quantifying the severity of a stroke using regression techniques should be explored. The developed models should also be trained and tested on alternative datasets or new data should be collected for this task.

\section{Acknowledgements}
This research was conducted independently by the authors. Both authors are affiliated with both Harvard Medical School and Thomas Jefferson High School for Science and Technology, though this research was conducted independently. All authors contributed equally.

\normalsize
\bibliography{references}


\end{document}